\documentclass[10pt,conference]{IEEEtran}
\IEEEoverridecommandlockouts
\AtBeginDocument{%
  }

\usepackage{tikz}
\usepackage{amsmath}
\usepackage{wasysym}
\usepackage{graphicx}
\usepackage{url} 
\usepackage{multirow}
\usepackage{booktabs}
\usepackage{amsmath}
\usepackage{amssymb}
\usepackage{tikz}
\usepackage{listings}
\usepackage{listings}
\usepackage[ruled,linesnumbered]{algorithm2e}
\usepackage{xcolor}
\SetCommentSty{mycommfont}

\SetKwComment{tcp}{{\color{blue}$\triangleright$~}}{}
\newcommand{\model}{{\textsf{Veritas}}}

\newcommand{\diff}[1]{\textcolor{black}{#1}}
\newcommand{\gensym}{\scalebox{0.85}{\Circle}}
\newcommand{\partsym}{\scalebox{0.85}{\LEFTcircle}}
\newcommand{\fullsym}{\scalebox{0.85}{\CIRCLE}}

\begin{document}

%\title{\model: A Semantic-Grounded Agentic Framework for Binary Memory Corruption Vulnerability Detection}

\title{\model: Grounding LLM Agents for Reliable Vulnerability Reasoning over Stripped Binaries
}

\makeatletter
\newcommand{\linebreakand}{%
  \end{@IEEEauthorhalign}
  \hfill\mbox{}\par
  \mbox{}\hfill\begin{@IEEEauthorhalign}
}
\makeatother

\author{
\IEEEauthorblockN{Xinran Zheng}
\IEEEauthorblockA{
University College London\\
xinran.zheng.23@ucl.ac.uk}
\and
\IEEEauthorblockN{Alfredo Pesoli}
\IEEEauthorblockA{
BynarIO\\
alfredo@bynar.io}
\and
\IEEEauthorblockN{Marco Valleri}
\IEEEauthorblockA{
BynarIO\\
marco@bynar.io}
\linebreakand
\IEEEauthorblockN{Suman Jana}
\IEEEauthorblockA{
Columbia University\\
suman@cs.columbia.edu}
\and
\IEEEauthorblockN{Lorenzo Cavallaro}
\IEEEauthorblockA{
University College London / BynarIO\\
\{l.cavallaro,lorenzo\}@\{ucl.ac.uk,bynar.io\}}
}
% \affiliation{%

% \affiliation{%

% \affiliation{%

% \affiliation{%

% \affiliation{%


\maketitle
\begin{abstract}
\diff{Frontier LLM agents show promise for vulnerability reasoning and can often localize suspicious code. However, access to program artifacts does not itself specify which facts an agent should carry forward to justify a vulnerability claim. Without an explicit obligation to justify each vulnerability claim against the evidence that decides it, safety-relevant facts may be present in the artifact yet absent from the reasoning that supports the claim. This creates a semantic gap between the facts available in the artifact and the evidence the agent actually uses to support the claim. In stripped-binary analysis, this gap is especially acute because source-level cues are removed and evidence is fragmented across noisy lifted IR and lossy decompiled views. Closing it requires grounding reasoning in recovered program semantics and checking the resulting claims against executable behavior.}

% Frontier LLM agents are increasingly used for vulnerability research, but their success often relies on semantically coherent program artifacts such as source code, identifiers, and structured program context. Stripped binaries break this assumption. Compilation and stripping remove much of the semantic structure needed for vulnerability reasoning, while lifting and decompilation recover only partial and sometimes inconsistent views. As a result, reliable vulnerability reasoning over stripped binaries requires more than exposing an agent to decompiled code or binary analysis tools: it calls for grounding agent reasoning in recovered program semantics and checking vulnerability hypotheses against executable behavior.

\diff{Building on this principle, we formulate binary vulnerability reasoning as a semantic grounding problem and present \model{}, a three-stage framework for reliable analysis over stripped binaries: a static-analysis \textit{Slicer} recovers witness-backed source-to-sink flows from lifted LLVM IR, an LLM-based \textit{Discover} stage aligns decompiled code with IR witnesses to construct vulnerability claims, and a multi-agent \textit{Validator} checks these claims through guided debugging and runtime oracles. Together, these stages turn fragmented binary views into checkable claims rather than direct agent inference.
We instantiate \model{} for out-of-bounds vulnerabilities and evaluate it on a curated benchmark with flow-level annotations. \model{} achieves 90\% recall, outperforms static, dynamic, binary-analysis, and agentic baselines, and reports no false positives among 623 exhaustively validated candidates and only two observed false positives in sampled audits. In a real-world case study, \model{} discovered a previously unknown Apple vulnerability that was confirmed and assigned a CVE, showing that grounded reasoning can produce actionable findings beyond the curated benchmark.}

\end{abstract}


\begin{IEEEkeywords}
Vulnerability Reasoning, LLM Agents, Stripped Binaries, Semantic Grounding
\end{IEEEkeywords}




\section{Introduction}
Automated vulnerability reasoning has advanced rapidly in source-level settings through static analysis, symbolic execution, and LLM-based code auditing~\cite{ding2024vulnerability,steenhoek2024err,li2024iris,guo2025repoaudit,guo2025bugscope}. These methods benefit from rich program evidence, including identifiers, types, control/data-flow structure, path constraints, and API-level cues. However, recent studies show that LLMs remain brittle, relying on shallow code signals, missing subtle semantic distinctions, or producing unfaithful explanations~\cite{ding2024vulnerability,steenhoek2024err,ullah2024llms,weissberg2025llm}. This challenge becomes sharper for deployed binaries: compilation and stripping remove source-level cues~\cite{pang2021sok,xu2005automatic}, so reasoning must reconstruct missing semantic evidence, such as object identity, propagation, bounds, and feasible triggering conditions, at real-world scale.

% When source code is available, analysts exploit identifiers, types, control flow, and data-flow cues to trace how external input reaches a security-sensitive operation. In practice, however, analysis often targets the compiled binaries that are actually deployed, which are frequently stripped of the debugging symbols, identifiers, and structural context that source-level reasoning relies on~\cite{pang2021sok,xu2005automatic}. Reasoning about such artifacts therefore requires reconstructing the semantic evidence that source-level analysis takes for granted, which demands deep expertise and does not scale to real-world binaries.

\begin{figure}[t]
    \centering
    \includegraphics[width=0.8\linewidth]{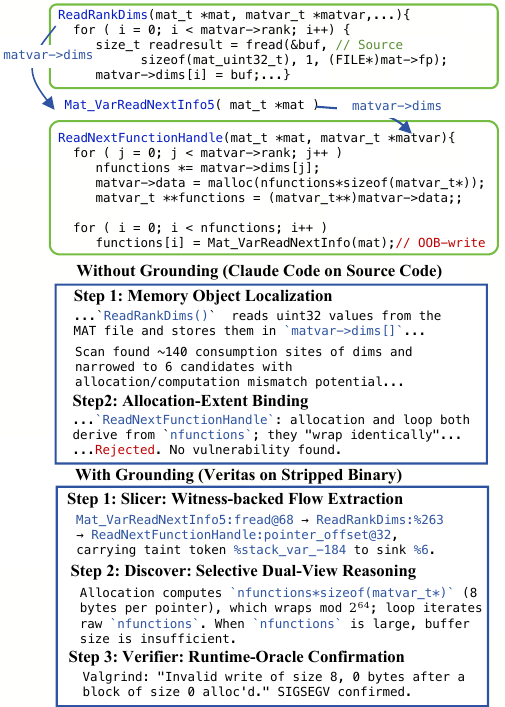}
    \caption{
    Failure case motivating semantic grounding. Even with source code, Claude Code with Opus 4.6 locates the sink but dismisses the bug by treating the allocation size and loop bound as equivalent, overlooking the \texttt{sizeof(matvar\_t*)} scaling factor. \model{} instead materializes the source-to-sink relation as a witness-backed flow, preserves the allocation-extent mismatch, and confirms the out-of-bounds write at runtime.
   % Failure case illustrating the need for semantic grounding. Even with source code, a frontier agent, Claude Code with Opus 4.6, locates the sink but dismisses the true vulnerability because it does not trace how the attacker-controlled value propagates through allocation and loop-bound computation. In contrast, \model{} operates on the stripped binary, traces the tainted value from source to sink through a witness-backed flow, identifies the allocation-extent mismatch, and confirms the out-of-bounds write at runtime.
    }
    \label{fig:intro}
\end{figure}

LLM agents provide a promising substrate for vulnerability analysis because they can summarize code, inspect candidate regions, and combine heterogeneous evidence across program representations~\cite{chen2024reasoning,liu2023harnessing,yang2025code,abramovich2024interactive,guo2025repoaudit,hussain2025vulbinllm}. Yet their effectiveness is bounded by the evidence carried into reasoning and the checks applied to their conclusions. Reliable reasoning requires more than locating suspicious operations: it must preserve the safety-deciding details, such as scaling factors, modular wraparound, object extents, and index-bound relationships. Existing agents do not do so reliably; even when such details are present in the code, they may be dropped before the claim is judged.
Figure~\ref{fig:intro} illustrates one such failure. Even with source code, a frontier agent locates the sink but dismisses the true bug by collapsing the \texttt{sizeof(matvar\_t*)} scaling factor into the abstraction that the allocation size and loop bound both derive from \texttt{nfunctions}. The evidence is present in source, yet the model discards the factor that breaks this equivalence. In binaries, where evidence is fragmented across lossy recovered views and not executable by construction, this failure mode becomes even more consequential.
% \diff{LLM agents provide a promising substrate for vulnerability reasoning~\cite{chen2024reasoning,liu2023harnessing,yang2025code,abramovich2024interactive,guo2025repoaudit,hussain2025vulbinllm}. They are often effective at local program comprehension, such as locating plausible risk operations or summarizing relevant code~\cite{lomshakov2024proconsul,chen2025locagent,liu2023harnessing}. However, current vulnerability-discovery ability may still rely on shallow code-level signals~\cite{weissberg2025llm}. Reliable vulnerability reasoning requires more than identifying suspicious regions: it depends on preserving the details, such as scaling factors, modular wraparound, or index-bound relationships, that determine safety. Agents do not make this choice reliably, and a deciding detail can be discarded even when it is present in the code. Figure~\ref{fig:intro} illustrates one such failure. Even with source code, a frontier agent locates the sink but dismisses the true bug, collapsing the \texttt{sizeof(matvar\_t*)} scaling factor into the abstraction that the allocation size and loop bound both derive from \texttt{nfunctions}. The evidence is present in plain source, yet the model discards the very factor that breaks this equivalence; such a collapse can suppress real vulnerabilities or, in its inverse form, produce confident unsupported reports, consistent with the low recall and high false-positive behavior observed in unconstrained agents~\cite{ding2024vulnerability,steenhoek2024err,ullah2024llms,yildiz2025benchmarking}.

This failure reflects a \textit{semantic gap}: safety-deciding evidence may be present in program artifacts but not preserved throughout the agent's reasoning. The gap is sharper in stripped binaries, where source-level cues are removed and decisive details are scattered across lossy views. Lifted IR preserves propagation and risk operations but is noisy and difficult to reason over end-to-end, while decompiled code is readable but often collapses object identity, bounds, and sink semantics into ambiguous offsets. Thus, even a recovered source-to-sink relation remains only a hypothesis unless execution confirms that the required path conditions, input values, object bounds, and failure condition hold together.
Reliable vulnerability reasoning therefore requires identifying the safety-deciding evidence for a candidate path, preserving it explicitly, and validating the claim against executable behavior.

Existing scaffolding mechanisms address parts of this problem, but leave evidence targeting implicit. Retrieval and summarization improve access to relevant code~\cite{guo2025bugscope,hussain2025vulbinllm,abramovich2024interactive}, yet still let the agent decide what to retain, so path-relevant facts may be summarized away. Call-graph and program-structure expansion provide stronger anchors by exposing interprocedural context~\cite{guo2025repoaudit,li2024iris,liu2023harnessing,nie2025vulnllm}, but these anchors remain coarse-grained and can miss dependencies through shared state, aliases, or indirect memory objects beyond direct caller-callee edges. Coverage-guided fuzzing explores executions, but without a target derived from the deciding evidence, it rarely reaches the specific path and memory state that trigger a corruption within a bounded budget. 
%These mechanisms therefore improve where the agent can look or what it can try, but they do not specify what evidence must be preserved and validated for a vulnerability claim. The central failure remains: the agent must still decide which details matter, exactly where unreliable vulnerability reasoning arises.
These mechanisms primarily improve evidence retrieval or execution exploration, but they do not explicitly specify which semantic evidence must be preserved to justify a vulnerability claim.

% Keep the opening pls
We argue that closing this gap requires \emph{semantic grounding}: explicitly selecting, materializing, and validating the semantic evidence that justifies a vulnerability claim. 
\textit{Static grounding} makes the evidence required by a vulnerability claim explicit before model reasoning, so the claim is not left to the agent's implicit choice of which details to retain. It constrains reasoning around the semantic relations that determine safety. \textit{Runtime grounding} then checks each claim against concrete execution. Static evidence establishes plausibility, not correctness. A candidate vulnerability remains a hypothesis until the required path conditions, object states, and failure conditions are simultaneously satisfied during execution. %since a statically plausible flow is only a hypothesis until its path conditions and failure state hold together, filtering confident but unsupported reports. 
Together, static and runtime grounding transform fragmented binary artifacts into executable, evidence-backed vulnerability claims.

% This paper investigates semantic grounding as an architectural principle for AI-assisted vulnerability reasoning over stripped binaries. Rather than asking whether frontier models can identify vulnerabilities, we ask how the semantic evidence required to justify vulnerability claims should be constructed, preserved, and validated when program semantics are fragmented across binary representations.

\diff{Based on this principle, we present \model{}, a semantically grounded framework for vulnerability reasoning over stripped binaries. In this work, we instantiate \model{} for out-of-bounds vulnerabilities, a prevalent class that stresses source-to-sink propagation, object-bound reasoning, and runtime feasibility. The grounding framework itself separates evidence construction, reasoning, and validation, providing a basis for investigating other vulnerability classes in future work.
\model{} combines a static-analysis \emph{Slicer}, an LLM-based \emph{Discover} stage, and an agentic \emph{Validator}. \emph{Slicer} recovers witness-backed flows from lifted LLVM IR, \emph{Discover} constructs candidate claims through selective dual-view reasoning over decompiled code and IR evidence, and \emph{Validator} confirms these claims through debugger-guided runtime checks.}

% The grounding architecture can be adapted to other vulnerability classes by changing the source/sink specifications and runtime oracles. \model{} combines a static-analysis \emph{Slicer} with agentic stages, \emph{Discover} and \emph{Validator}. \emph{Slicer} faithfully reconstructs witness-backed source-to-sink flows from lifted LLVM IR, preserving vulnerability-relevant propagation evidence that may be obscured in decompiled views. \emph{Discover} performs selective dual-view reasoning, using decompiled code for interpretable control and constraint reasoning while consulting IR witnesses only where object, bound, or sink semantics require higher fidelity. \emph{Validator} tests each candidate claim through a multi-agent workflow for guided debugging, breakpoint inspection, and runtime-oracle checks. Together, These stages combine semantic fidelity, controlled reasoning scope, and executable confirmation, turning fragmented binary representations into checkable vulnerability claims.}

\diff{In this paper, we make the following contributions:}
\begin{itemize}
    \item \diff{
    We characterize a semantic gap in AI-assisted vulnerability reasoning over stripped binaries, where safety-deciding evidence may be present in binary artifacts yet fail to be preserved throughout reasoning. We formulate vulnerability discovery as grounded claim construction, where claims are built from witness-backed propagation evidence and confirmed against executable behavior.}
    %
    %We identify a semantic gap in vulnerability reasoning over stripped binaries, where agents may locate suspicious regions but fail to preserve safety-deciding details. We formulate the task as grounded claim construction, where claims must be built from recovered propagation evidence and confirmed against executable behavior.}
    
    \item \diff{
    We propose a semantically grounded architecture for AI-assisted vulnerability reasoning over stripped binaries and instantiate it in \model{}. It explicitly separates evidence construction, claim construction, and claim validation through three cooperating stages: a static-analysis \textit{Slicer}, an LLM-based \textit{Discover} stage, and an agentic \textit{Validator}.}
    %We present \model{}, a semantically grounded framework for vulnerability reasoning over stripped binaries. It realizes a three-stage grounded architecture: a static-analysis \textit{Slicer} for witness-backed flow recovery, an LLM-based \textit{Discover} stage for selective dual-view claim construction, and an agentic \textit{Validator} for runtime confirmation through debugger-visible artifacts and concrete oracles.}

    \item
    We evaluate \model{} on real-world out-of-bounds vulnerabilities with verified flow-level ground truth. Veritas achieves 90\% recall, substantially outperforms representative static-analysis, binary-analysis, fuzzing, and agentic baselines, reports no false positives among 623 exhaustively validated candidates, and discovers a previously unknown vulnerability in software from the Apple ecosystem that was confirmed with a CVE.
    %We instantiate and evaluate \model~ on real-world out-of-bounds vulnerabilities in stripped binaries with verified flow-level ground truth. \model~achieves 90\% recall, outperforms static, dynamic, binary-analysis, and agentic baselines, reports no false positives among 623 exhaustively validated candidates, and discovers a previously unknown Apple vulnerability confirmed with a CVE.}
\end{itemize}

\section{Motivation}

\begin{figure*}[t]
    \centering
    \includegraphics[width=1.0\linewidth]{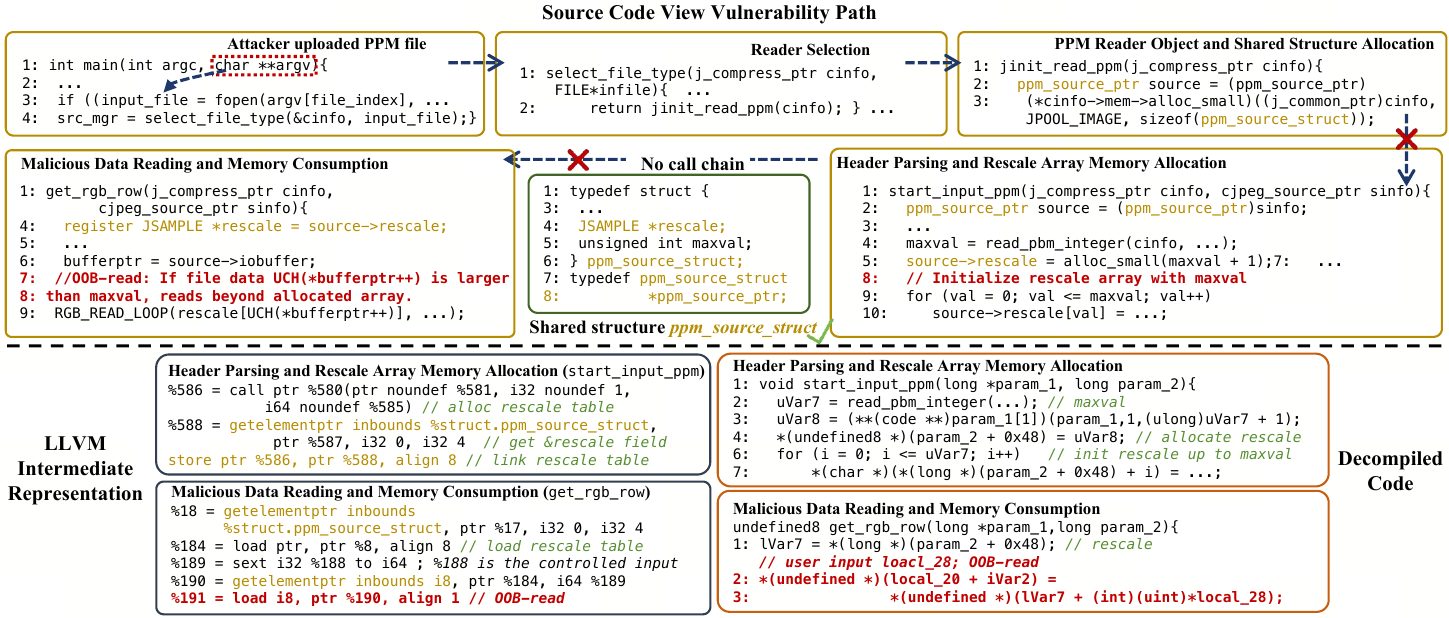}
    \caption{Realistic program artifacts from CVE-2020-13790, an out-of-bounds read in the PPM reader. The vulnerability arises from a shared-memory dependence, where \texttt{jinit\_read\_ppm()} allocates a \texttt{ppm\_source\_struct}, \texttt{start\_input\_ppm()} populates its rescale array, and \texttt{get\_rgb\_row()} later consumes this array using an attacker-controlled index without checking the relationship between \texttt{maxval} and \texttt{bufferptr}. All functions are connected through the global structure \texttt{ppm\_source\_struct}, not by direct call edges. For readability, LLVM IR snippets are shown using source-level LLVM IR.}
    \label{fig:motivation}
\end{figure*}

\subsection{Challenges in Reasoning Vulnerabilities in Binaries}
\diff{
To illustrate vulnerability reasoning over stripped binaries, we use CVE-2020-13790, an out-of-bounds read caused by inconsistent size assumptions about a shared lookup table. As shown in Figure~\ref{fig:motivation}, \texttt{jinit\_read\_ppm} allocates a \texttt{ppm\_source\_struct}, \texttt{start\_input\_ppm} initializes its \texttt{rescale} buffer using the parsed \texttt{maxval}, and \texttt{get\_rgb\_row} later indexes this buffer with attacker-controlled input. The bug is triggered when an input byte exceeds the \texttt{maxval} used to size the array. While a human analyst can connect these operations using source-level names and structure fields, stripped-binary reasoning must recover the same propagation, object-bound, and feasibility evidence from lifted, decompiled, and executable views. The following subsections detail three challenges that motivate \model{}.
% To illustrate challenges in vulnerability reasoning over stripped binaries, we use CVE-2020-13790, an out-of-bounds read caused by inconsistent size assumptions about a shared lookup table.
% This example is a concrete instance of the reasoning problem studied in this paper: a vulnerability claim depends on connecting input propagation, object-bound reasoning, and runtime feasibility checking from fragmented binary evidence.
% As shown in Figure~\ref{fig:motivation}, \texttt{jinit\_read\_ppm} allocates a \texttt{ppm\_source\_struct}, \texttt{start\_input\_ppm} initializes the \texttt{rescale} buffer using the parsed \texttt{maxval}, and \texttt{get\_rgb\_row} later indexes this buffer with attacker-controlled input.
% The vulnerable behavior occurs when an input byte exceeds the \texttt{maxval} used to size the array.
% A human analyst can connect these operations using source-level names and structure fields, but the same reasoning over stripped binaries requires recovering the relevant propagation, object-bound, and feasibility evidence from lifted, decompiled, and executable views. The following subsections detail three challenges that motivate grounding design in \model{}.
}

% a vulnerability claim is not localized to a single suspicious access, but depends on reconstructing how input-derived values propagate through shared state, how object bounds are established and later consumed, and whether the triggering condition is feasible at runtime. 

% of BMCV detection, we use CVE-2020-13790, an out-of-bounds read caused by inconsistent size assumptions about a shared lookup table, as a motivating example because it compactly captures key difficulties of BMCV. As shown in the source-code view of Figure~\ref{fig:motivation}, the vulnerability is not localized to a single block of code, but spans a disjoint lifecycle involving the global structure \texttt{ppm\_source\_struct}: \texttt{jinit\_read\_ppm} allocates the structure object, \texttt{start\_input\_ppm} initializes the \texttt{rescale} buffer according to the parsed \texttt{maxval}, and \texttt{get\_rgb\_row} later consumes this buffer, triggering an out-of-bounds access when an attacker-controlled input byte exceeds the \texttt{maxval} used to size the array. While a human expert may connect these functions through source-level names and structure fields, automating the same reasoning in stripped binaries requires recovering such relations from lossy lifted and decompiled representations. The following subsections detail three challenges that motivate semantic grounding in \model.

\subsubsection{Propagation beyond Call Chains}
\diff{Existing approaches often rely on call-chain reasoning to approximate vulnerability propagation~\cite{guo2025repoaudit,li2024iris,liu2023harnessing}; however, this is insufficient when the relevant coupling is defined by shared program state rather than a single function or direct caller-callee chain. As illustrated in Figure~\ref{fig:motivation}, \texttt{jinit\_read\_ppm()}, \texttt{start\_input\_ppm()}, and \texttt{get\_rgb\_row()} share the global \texttt{ppm\_source\_struct} with no direct call edges, so allocation, bound initialization, and the out-of-bounds read occur in disjoint locations. Connecting them requires resolving aliases and reconstructing field-level uses of memory objects, which call-chain context alone does not capture.}

\subsubsection{Fragmented Semantics across Binary Views}
\label{sec: semantic loss}
\diff{Reasoning over shared program state requires understanding how vulnerability-relevant objects are created, constrained, propagated, and consumed across different execution stages. In stripped binaries, this evidence is fragmented across recovered representations. As Figure~\ref{fig:motivation} illustrates, lifted LLVM IR preserves low-level propagation and memory-operation facts through instructions such as \texttt{getelementptr} and explicit loads, stores, and pointer arithmetic. However, it is saturated with transient SSA variables and compiler-introduced bookkeeping, which inflate context size and make vulnerability logic difficult for an LLM agent to interpret. Decompiled code provides a more concise view for control-flow and constraint reasoning, but it collapses memory operations into generic offset expressions, such as \texttt{param\_2 + 0x48}, obscuring object correspondence, array bounds, and sink semantics. The evidence needed for vulnerability reasoning is thus split across individually incomplete representations. We call this the precision-interpretability dilemma: object identity and sink semantics are explicit only in IR, while interpretable structure is clearer only in decompiled code.}

\subsubsection{Executable Feasibility Verification}
\diff{A vulnerability claim is incomplete unless the inferred behavior can be tied to a feasible execution: even a plausible source-to-sink flow may not establish whether the required branch decisions, input values, and object bounds can hold simultaneously. As illustrated in Figure~\ref{fig:motivation}, the out-of-bounds read fires only when the PPM parser selects the relevant decoding path and an input byte exceeds the parsed \texttt{maxval}. A suspicious flow therefore remains a hypothesis until its path constraints and failure condition are grounded in concrete execution.
% As illustrated in Figure~\ref{fig:motivation}, the out-of-bounds read is triggered only when the PPM parser selects the relevant decoding path and an input byte exceeds the parsed \texttt{maxval}. Thus, the claim depends not only on connecting the bound definition in \texttt{start\_input\_ppm} to the access in \texttt{get\_rgb\_row}, but also on checking whether the triggering condition can occur at runtime. A suspicious flow therefore remains a hypothesis unless its path constraints and failure condition are grounded in concrete execution behavior.
}

These challenges share a common root: over stripped binaries, the evidence needed for a vulnerability claim is fragmented across lossy views and not executable by construction. Addressing them requires semantic grounding: constraining reasoning with recovered source-to-sink evidence and checking resulting claims against execution. We realize this principle in \model; the next section details its design.

\section{Methodology}
\subsection{Problem Definition}
\diff{We study vulnerability reasoning over stripped binaries. Given a stripped executable, the task is to discover and verify vulnerability claims without source-code access. Each claim consists of an attacker-controlled propagation chain to a vulnerability-relevant operation, the violated semantic condition, and runtime evidence supporting the violation. We instantiate this task on out-of-bounds vulnerabilities, focusing on memory operations whose safety depends on object bounds.}

\begin{figure*}[t]
    \centering
    \includegraphics[width=0.9\linewidth]{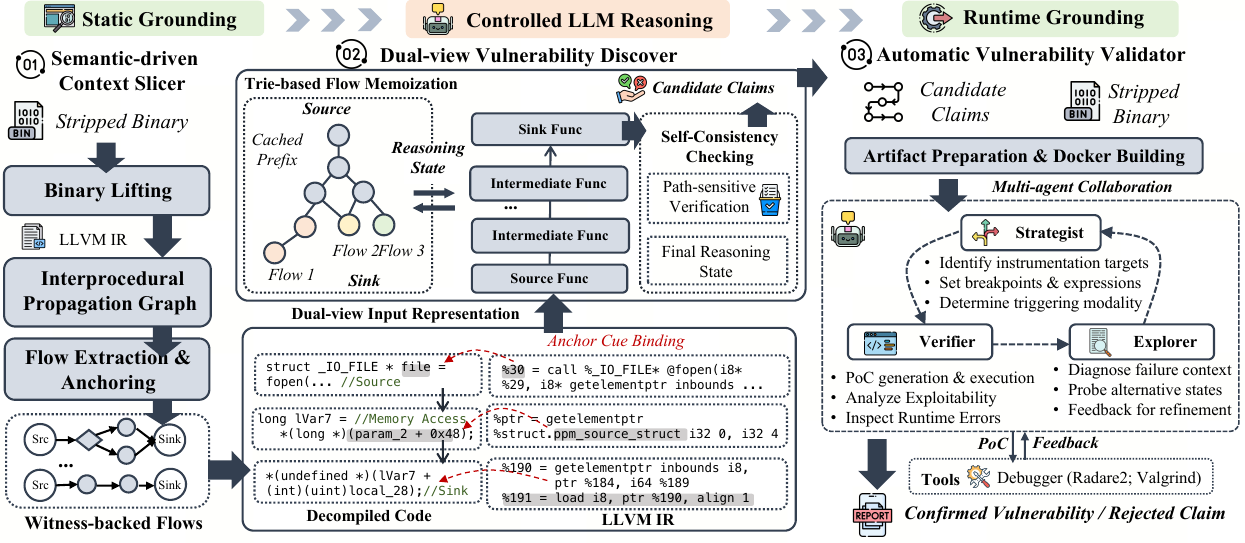}
    \caption{The workflow of \model{}, organized into static grounding, controlled LLM reasoning, and runtime grounding. (1) The \textit{Slicer} extracts witness-backed flows with provenance-relevant anchor cues from stripped binaries. (2) \textit{Discover} reasons step-wise along these flows, selectively consulting LLVM IR when decompiled code is ambiguous. (3) The \textit{Validator} confirms candidate claims through debugger-guided execution and runtime oracles.}
    \label{fig:arch}
\end{figure*}

\subsection{Overview of Framework}
Building on the design insights above, \model{} organizes vulnerability reasoning over stripped binaries into three stages spanning two grounding layers, as illustrated in Figure~\ref{fig:arch}. The \textbf{Semantic-driven Context Slicer} provides static grounding, recovering witness-backed source-to-sink flows that determine what evidence the model reasons over. The \textbf{Dual-view Vulnerability Discover} then performs controlled LLM reasoning along each flow to construct candidate claims. Finally, the \textbf{Automatic Vulnerability Validator} provides runtime grounding, confirming these claims against executable behavior. This separation keeps evidence recovery, claim construction, and executable confirmation distinct, turning fragmented binary representations into checkable vulnerability claims rather than unconstrained model speculation.

\subsection{Semantic-driven Context Slicer}
The first stage of \model\ provides static grounding: the \textit{Slicer} recovers compact, evidence-backed source-to-sink flows from lossy binary representations, so that the downstream \textit{Discover} stage reasons over explicit propagation evidence rather than unconstrained decompiled code. Operating before any LLM invocation, it takes a stripped binary and source/sink specifications and produces a set of grounded evidence flows ($\Pi=\{\pi_1,\ldots,\pi_M\}$). Each flow ($\pi$) contains an ordered sequence of recovered functions ($(f_1,\ldots,f_N)$), a verified propagation witness from a source-derived token to a sink, and compact anchor cues indicating which value, object, bound expression, or sink operand to track at each function. In our out-of-bounds instantiation, sources are attacker-controlled input operations and sinks are bound-dependent memory operations; other classes instantiate the same \textit{Slicer} by changing the source/sink specifications and witness acceptance rules.

% A fundamental challenge in binary vulnerability detection lies in the extreme imbalance between program size and vulnerability-relevant semantics: only a small fraction of instructions participate in exploitable behaviors, while the majority constitute benign control logic. Directly feeding whole functions or binaries into LLMs, therefore, introduces substantial noise and quickly exceeds practical context budgets~\cite{liu2024lost}. To address this challenge, we design a Semantic-driven Context Slicer that operates entirely without LLMs and extracts compact, vulnerability-relevant context from stripped binaries. The \textit{Slicer} takes as input a stripped binary $\mathcal{B}$ and produces a set of candidate vulnerability flows $\Pi=\{\pi_1,\ldots,\pi_M\}$. Each flow object $\pi$ contains an ordered sequence of recovered functions $(f_1,\ldots,f_N)$ connecting an attacker-controlled input to a security-sensitive memory operation, together with compact grounded labels derived from the underlying propagation of the taint. %and dedicated endpoint annotations used by the downstream Detector. %the corresponding taint sets $(T_1,\ldots,T_N)$ carried along the flow.

\subsubsection{Interprocedural Propagation Graph Construction} 
\label{sec:slicer:vfg}
Vulnerability reasoning depends on how source-derived values, objects, and shared state propagate across functions before reaching a security-relevant operation. To recover such evidence from a stripped binary, the \textit{Slicer} lifts binaries with RetDec~\cite{retdec} into LLVM IR and extracts an interprocedural fact base $\mathcal{P}$ recording per-function instructions, SSA def-use chains, call and global-access relations, return-value propagation, pointer/arithmetic operations, and sink-related facts.
From $\mathcal{P}$, the \textit{Slicer} materializes a typed interprocedural propagation graph
$\mathcal{G}_{\mathcal{P}} = (\mathcal{F}, \mathcal{E}, \kappa, \mu)$,
where $\mathcal{F}$ is the set of recovered functions, $\mathcal{E} \subseteq \mathcal{F} \times \mathcal{F}$ is the set of propagation edges, $\kappa : \mathcal{E} \rightarrow {\mathtt{call}, \mathtt{return}, \mathtt{global}}$ assigns each edge a propagation kind, and $\mu$ stores the metadata needed to transfer source-derived provenance across the edge, such as actual-to-formal mappings, return-value mappings, or the global object involved in a write-read dependency.

This graph defines the interprocedural search space for source-to-sink evidence paths. It is more informative than a call graph because it captures propagation through calls, returns, and shared state, but it does not by itself prove a vulnerability or even a complete source-to-sink flow. The latter is checked by the witness extraction step. RetDec also produces decompiled C for the same recovered functions; \model\ pairs LLVM IR and decompiled code at function granularity so that the downstream \textit{Discover} stage can use IR for precise propagation evidence and decompiled code for higher-level control, object, and constraint reasoning.

\subsubsection{Witness-based Flow Extraction}
\label{sec:extraction}
The graph $\mathcal{G}_{\mathcal{P}}$ defines the interprocedural search space, but a reachable source and sink may not carry the same value, object, or constraint. The \textit{Slicer} therefore enumerates bounded candidate paths and accepts only those supported by a successful provenance-transfer trace. Let $\mathcal{F}_{src} \subseteq \mathcal{F}$ and $\mathcal{F}_{sink} \subseteq \mathcal{F}$ denote functions containing source and sink operations for the target class. For each $f_0 \in \mathcal{F}_{src}$, the \textit{Slicer} initializes source-derived provenance tokens from LLVM-IR facts and considers bounded paths
$p: f_0 \xrightarrow{e_0} f_1 \cdots \xrightarrow{e_{k-1}} f_k$,
where $f_k \in \mathcal{F}_{sink}$. The checker validates each path step by step: within each function, it expands the token set through local def-use chains, load/store links, aliases, and pointer or arithmetic operations; across each edge $e_i=(f_i,f{i+1})$, it transfers tokens according to $\kappa(e_i)$ and $\mu(e_i)$, such as call-argument mappings, return-value mappings, or global write-read dependencies. If no provenance token can be transferred at any step, the path is rejected.
% Let $\mathcal{F}{src} \subseteq \mathcal{F}$ and $\mathcal{F}{sink} \subseteq \mathcal{F}$ denote functions containing source and sink operations specified by the target vulnerability class. For each $f_0 \in \mathcal{F}{src}$, the \textit{Slicer} initializes source-derived provenance tokens from LLVM-IR facts and considers bounded paths
% $p: f_0 \xrightarrow{e_0} f_1 \cdots \xrightarrow{e{k-1}} f_k$,
% where $f_k \in \mathcal{F}{sink}$. The checker validates each path step by step: within each function, it expands the token set through local def-use chains, load/store links, aliases, pointer and arithmetic operations, and lifted-IR memory-staging patterns; across each edge $e_i=(f_i,f{i+1})$, it transfers tokens according to $\kappa(e_i)$ and $\mu(e_i)$, such as call argument mappings, return-value mappings, or global write-read dependencies. If no provenance token can be transferred at any step, the path is rejected.

A path is accepted as a propagation witness only when a source-derived token reaches a class-specific sink. The \textit{Slicer} then constructs a grounded evidence flow $\pi$ from the accepted witness and adds only helper frames needed to interpret the flow, such as functions that receive the propagated token, compute relevant bounds, or forward return values, each justified by SSA provenance, field or global tags, or equivalent propagation evidence. The resulting flow is a compact reasoning unit, containing the source, sink, ordered frames, per-function provenance tokens, and the underlying witness.
\subsubsection{Witness Slice Structure}
Each grounded flow is represented as a witness slice, i.e., a function-level projection of the accepted witness with evidence-supported helper frames. For every retained function $f_n$, the \textit{Slicer} attaches a compact grounded label $\ell_n$ identifying the provenance-relevant value, shared object tag, bound-related expression, or sink operand to track in that function. Source and sink functions additionally receive endpoint annotations recording concrete IR-supported source/sink tokens and callsite information. The \textit{Slicer} also assigns each function a provenance class indicating how it is retained: on the accepted witness, through shared-state (\texttt{global\_proven}) or local def-use (\texttt{defuse\_proven}) evidence, or only for interpretive continuity (\texttt{context\_only}).

Together, the grounded label, endpoint annotations, and provenance class form the anchor cues $\mathcal{A}(f_n)$ for each retained function. These cues define the per-function static grounding interface to \textit{Discover}, specifying what to track and how strongly the function is supported by the witness or auxiliary propagation evidence. A binary may yield multiple grounded flows, each passed to \textit{Discover} for claim construction.

\subsection{Dual-view Vulnerability Discover}
\label{sec:dual_view_discovery}
Given the grounded flows $\Pi$ produced by the \textit{Slicer}, \textit{Discover} performs controlled, function-by-function LLM reasoning along each witness-backed flow. Rather than free-form agentic retrieval, it reasons only over the flow-local views the \textit{Slicer} emits, constructing a candidate claim when the flow supports a class-specific safety violation. Candidate claims are then forwarded to the \textit{Validator}.
% Given the candidate flows $\Pi$ produced by the \textit{Slicer}, the \textit{Detector} performs step-wise reasoning over statically grounded, witness-backed traces. The LLM is not responsible for discovering taint propagation from the whole binary by itself; instead, propagation is precomputed by the \textit{Slicer} and provided as a fixed interprocedural context. The \textit{Detector} analyzes this context to recover the missing reasoning dimensions, including control-flow feasibility, object correspondence, and arithmetic bound constraints, to determine whether the flow is vulnerability-relevant. Flows judged as potentially vulnerable are then forwarded to the \textit{Validator} for dynamic confirmation. Algorithm~\ref{alg:detector} summarizes the procedure.

\subsubsection{Dual-view Input Representation.}
\label{sec:dual_view}
Let $D(f_t)$ denote the decompiled code of function $f_t$ and $I(f_t)$ its lifted LLVM IR, both produced by RetDec. Following the precision-interpretability dilemma of Section~\ref{sec: semantic loss}, \textit{Discover} adapts each function's representation to its role in the flow, using the anchor cues $\mathcal{A}(f_t)$ to identify the tokens and operations to track: decompiled code is compact for control-flow and constraint reasoning but collapses memory dependencies into ambiguous offset expressions, while lifted IR preserves low-level provenance but is too verbose for full-flow reasoning. \textit{Discover} therefore consults both views at source and sink functions, where provenance alignment and sink semantics require higher fidelity, and the decompiled view alone at intermediate functions. Formally, the function view is:
\begin{equation}
\mathcal{R}(f_t)=
\begin{cases}
\langle I(f_t),\, D(f_t)\rangle, & f_t \in \mathcal{F}_{src} \cup \mathcal{F}_{sink},\\
D(f_t), & \text{otherwise}.
\end{cases}
\end{equation}
When dual-view input is used, the prompt presents both representations together with the endpoint annotations in $\mathcal{A}(f_t)$. For intermediate functions, the grounded label in $\mathcal{A}(f_t)$ specifies what to track, while the provenance class indicates how strongly the function is supported by static evidence.
% When dual-view is triggered, the prompt presents both representations together with the relevant endpoint annotations from $\mathcal{A}(f_t)$. At intermediate functions, the \textit{Detector} reasons over $D(f_t)$ alone; the propagation tokens in $\mathcal{A}(f_t)$ identify which value to track, while the provenance class indicates the evidential strength of the function's presence in the flow.

\subsubsection{Step-wise Reasoning}
Instead of analyzing an entire flow in one LLM invocation, \textit{Discover} processes each grounded flow function by function. This keeps the model's context local to the witness-backed flow and avoids the loss of fine-grained attention caused by long mixed IR/decompiled inputs~\cite{liu2024lost}. Throughout, \textit{Discover} maintains a running semantic inventory of propagated values, object correspondences, and the class-specific constraints needed to interpret the sink.
% Analyzing an entire flow in a single LLM invocation is impractical: the combined IR and decompiled code of all functions can exceed context limits, and fine-grained semantic attention degrades with input length~\cite{liu2024lost}. We therefore analyze each flow one function at a time, carrying forward the reasoning state through the accumulated prompt-response history of the current flow prefix and a running buffer and bounds inventory. This inventory records propagated values, memory-object aliases, and bound-related constraints needed for sink verification.

Let $\mathcal{S}_n$ denote the accumulated reasoning state after processing prefix $(f_1,\ldots,f_n)$: the prompt-response history and running inventory, not a separate static-analysis object. At each step, the prompt updates the state only when the current function adds, refines, or invalidates prior facts. With $\Phi_{\theta}$ the step-wise LLM reasoning function, the update at step $n$ is:
\begin{equation}
  \mathcal{S}_n \leftarrow \Phi_{\theta}(\mathcal{S}_{n-1},\; \mathcal{R}(f_n),\; \mathcal{A}(f_n)), \quad n = 1, \ldots, N.
  \label{eq:step}
\end{equation}
where $\mathcal{R}(f_n)$ is the role-dependent function view and $\mathcal{A}(f_n)$ provides the grounded label, provenance class, and, for source or sink functions, endpoint annotations.

The prompts are specialized by function role. At the source step, \textit{Discover} aligns the source provenance token with the decompiled view and initializes the inventory. At intermediate steps, it updates the inventory through assignments, arithmetic, aliases, and control-flow conditions affecting propagation. At the sink step, it aligns the sink label with the corresponding operation and checks whether accumulated evidence supports a class-specific safety violation.

% The prompts are specialized according to the function's role in the flow. At the source step, the \textit{Detector} maps the IR taint label into the decompiled view and initializes the inventory of propagated values and bound-related facts. Intermediate steps incrementally update this inventory through assignments, arithmetic, and control-flow decisions that affect feasible propagation. At the sink step, the \textit{Detector} maps the sink label to the corresponding operation and performs end-to-end path-sensitive verification, including branch feasibility and bound consistency.

\subsubsection{Trie-based Flow Memoization}
\label{sec:trie_memoization}
Many grounded flows share prefixes, especially when they originate from the same parsing logic or reach nearby sinks, so recomputing each flow's reasoning state would cause redundant LLM invocations. However, reuse must distinguish flows that traverse the same functions but track different provenance-relevant values or sink operands. \model\ memoizes prefixes using the compact representation passed from the \textit{Slicer} to \textit{Discover}. Each retained function contributes a key element $q_i=(\tilde{f}_i,\ell_i)$, where $\tilde{f}_i$ is the recovered function identifier in the emitted flow, and $\ell_i$ is the compact grounded label attached to that function. In implementation, $q_i$ is serialized as \texttt{fn:@}; the \texttt{data} field identifies the value or operand to track, but is not the full internal provenance state maintained by the \textit{Slicer}. The cache $\mathcal{C}$ is a prefix trie over these key elements. A root-to-depth path $n$ represents:
\begin{equation}
\pi^{(n)}=(q_1,\ldots,q_n)=((\tilde{f}_1,\ell_1),\ldots,(\tilde{f}_n,\ell_n)), 
\end{equation}
and stores the reasoning state $\mathcal{S}_n$ after that prefix. For a new flow, \textit{Discover} resumes from the longest exactly matched prefix and invokes the LLM only on the remaining suffix. 

\subsubsection{Self-Consistency Checking and Claim Construction}
\label{sec:detector:decision}
After reaching the sink function $f_N$, \textit{Discover} does not treat the final semantic state $\mathcal{S}_N$ as a verdict. Instead, it performs a self-consistency check over the grounded flow $\pi=(f_1,\ldots,f_N)$ using the accumulated semantic-state sequence $(\mathcal{S}_1,\ldots,\mathcal{S}_N)$, combining path-sensitive verification with a sink-condition check. The path-sensitive verification confirms that the accessed value remains transitively derived from the root provenance label along a feasible chain, rather than being sanitized, overwritten, or gated by infeasible branches at intermediate steps. \textit{Discover} then reconstructs the sink safety condition from the accumulated semantic inventory, such as object bounds, allocation sizes, index expressions, and effective access ranges, and checks whether it is violated. A candidate claim is emitted only when both hold. This decision stage prioritizes recall: plausible claims are preserved for downstream confirmation, while final confirmation and false-positive filtering are deferred to the \textit{Validator}. Multiple sink accesses within the same flow that correspond to the same candidate violation are merged into a single report entry.

\subsection{Automatic Vulnerability Validator}
\textit{Discover} produces a vulnerability hypothesis over statically grounded flow objects, but this hypothesis must still be tied to executable behavior. The \textit{Validator} therefore serves as a second grounding stage: rather than asking an LLM to judge a candidate in isolation, it translates \textit{Discover} outputs into debugger-visible artifacts and confirms them through auditable tool interactions over the target binary.

\subsubsection{Artifact Preparation}
Before validation, \model{} aligns each vulnerability claim over RetDec-recovered functions with debugger-visible symbols, addresses, and code views. The target binary is analyzed with Radare2~\cite{radare2} to materialize structured pseudo-code (\texttt{pddj}) for slice functions; RetDec function names and addresses are mapped to Radare2 functions, and RVAs are computed for reported and sink locations. When the two views diverge, \textit{Discover} explanations are rewritten into the Radare2/r2dec view so that breakpoints, watched expressions, and memory inspections refer to the same runtime objects. The resulting JSON artifacts (mapped functions, RVAs, pseudo code, claims, and flow context) give the validation agents a consistent executable grounding interface.

\subsubsection{Automatic Framework}
Using the prepared artifacts, the \textit{Validator} orchestrates three agents, \textit{Strategist}, \textit{Verifier}, and \textit{Explorer}, to test each hypothesis against runtime evidence. The \textit{Validator} is not merely a false-positive filter; it is a runtime grounding layer that constrains reasoning through executable artifacts, debugger observations, and memory-checking oracles. A candidate is validated only when evidence supports the reported sink, access pattern, and root cause.

\emph{Strategist:} It converts \textit{Discover} claims into executable validation plans. Given the \textit{Discover} report, witness-backed flow, mapped artifacts, and target binary, it resolves each claim into runtime-checkable targets, such as the sink location, relevant symbol, memory region, and input channel. It then selects instrumentation points, breakpoints, triggering modality, and confirmation criteria. When validation fails, it uses feedback from the \textit{Verifier} and \textit{Explorer} to revise the plan and avoid redundant probes. Its output specifies how to exercise the binary and what runtime evidence to collect.
% It converts vulnerability claims from \textit{Discover} into an executable validation plan.
% It takes as input a report of \textit{Discover}, the witness-backed flow object, the mapped artifact layer, and the target binary. From these inputs, it resolves claims into runtime-checkable targets, including the sink location, relevant symbol, memory region, and input channel. It then selects functions and program points to instrument, translates them into breakpoints and observations, chooses the triggering modality (e.g., file, command line, network), and defines the conditions for confirming or refuting the hypothesis. When validation attempts fail, the \textit{Strategist} incorporates feedback from the \textit{Verifier} and \textit{Explorer} to revise the plan and avoid redundant probes. The output is a structured plan specifying how to exercise the binary and what runtime evidence to collect.

\emph{Verifier:} It executes the plan and collects runtime evidence. Guided by the \textit{Strategist}'s plan, it prepares the input channel, generates or adapts PoCs, and runs the target under debugging tools with controlled breakpoints. Its goal is not merely to trigger a crash, but to determine whether the observed runtime behavior supports the reported sink, access pattern, and root cause. If execution produces a segmentation fault, the \textit{Verifier} checks whether the crash location and trace match the detector hypothesis. When no explicit crash occurs, it invokes Valgrind~\cite{valgrind} to inspect invalid accesses and object-lifetime violations. When Valgrind reports no errors, e.g., stack-based off-by-one bugs, the \textit{Verifier} relies on targeted breakpoints and memory inspection to check whether the relevant memory region is modified or accessed consistently with claims.

\emph{Explorer:} It is activated when the \textit{Verifier} cannot obtain sufficient evidence from the current plan. It does not change the vulnerability hypothesis under test; instead, it reruns the same PoC with supplementary breakpoints, symbols, and memory observations to search for nearby evidence. This may extend beyond the original slice to functions related to branch conditions, input sanitization, or intermediate states needed to trigger the bug. The results are returned to the \textit{Strategist} to refine the next validation plan.

%The resulting observations are returned to the \textit{Strategist} to refine the next validation plan.

Together, the three agents operate a reasoning, execution, and feedback loop until a vulnerability is confirmed, ruled out, or the system reaches the maximum iterations to control cost and avoid endless exploration.

\section{Evaluation}
We conducted comprehensive experiments to evaluate \model\ and answer the following research questions:

\textbf{RQ1.} How effectively does \model{} identify and validate grounded vulnerability claims from stripped binaries compared with existing analysis and agentic auditing approaches?

\textbf{RQ2.} How does \model\ compare with agentic baselines in cost, and where are its main bottlenecks?

\textbf{RQ3.} How do the grounding components contribute to the effectiveness of vulnerability reasoning?

%How do the individual components individually contribute to detection effectiveness?

% To answer these questions, we conduct a comprehensive evaluation on a curated dataset of real-world binaries and compare \model\ against representative baselines, report efficiency metrics, and perform targeted ablation studies.

% RQ1: effectiveness: FP (after filtering) TP $\rightarrow$ Repoaudit; show results on each CWE + average cost (Time and Financial) with different LLM models (GPT-5 \& Claude?)

% RQ2: unknown vulnerability? LLM unknown/zero-day $\rightarrow$ latest project/after knowledge cutoff?

% RQ3: quality of each module: Slicing: correctness of dangerous flow (coverage of dangerous flows); Detector: FP \& TP before; Validator: FP \& TP after filtering; 
\subsection{Experiment Setup}
We implement \model\ as a modular pipeline combining deterministic static analysis, LLM-based reasoning, and runtime validation. RetDec~\cite{retdec} lifts binaries into LLVM IR and decompiled C. The \textit{Slicer} is implemented as \textit{defuse}, a custom LLVM-IR analysis framework, with bounded witness verification backed by a Rust CompactIR verifier. \textit{Discover} runs on GPT-5.4~\cite{openai_gpt54_2026} over the emitted flow objects, as a single-model stage rather than multi-agent. The \textit{Validator} is a multi-agent runtime-confirmation stage built on AutoGen~\cite{wu2024autogen}, Radare2~\cite{radare2}, and Valgrind~\cite{valgrind}.
% We implement \model\ as a modular analysis pipeline that integrates deterministic static analysis with an LLM-driven agentic framework. The system is composed of several specialized components, each responsible for a distinct stage of vulnerability analysis. For static analysis, we employ RetDec~\cite{retdec} as the primary backend for binary lifting and decompilation, producing LLVM IR that enables SVF-Engine~\cite{svf} to perform precise value-flow analysis. Based on the value-flow graph, the \textit{Slicer} conducts backwards and forward value-flow slicing to track shared memory states. For agentic analysis, the overall reasoning and coordination logic is implemented using  AutoGen~\cite{wu2024autogen}, with GPT-5.4~\cite{openai_gpt54_2026} serving as the underlying LLM to orchestrate tool invocation, intermediate result interpretation, and analysis decision-making. \textit{Detector} analyzes candidates from \textit{Slicer} to infer potential memory-corruption patterns. The \textit{Validator} then leverages Radare2~\cite{radare2} for interactive debugging and runtime memory checking. Valgrind~\cite{valgrind} is also used for memory checking for specific vulnerability types. 

\subsection{Dataset and Ground Truth}
\label{sec:dataset}
Existing vulnerability datasets~\cite{redini2020karonte,juliet-ccpp-13} typically label vulnerable programs, commits, or crashing inputs, but lack the flow-level evidence needed to adjudicate reported claims over stripped binaries. We therefore construct a curated benchmark of real-world out-of-bounds vulnerabilities with manually verified source-to-sink annotations, collected from SEC-Bench~\cite{lee2025sec}, ARVO~\cite{mei2024arvo}, and official CVE-bearing open-source repositories. Each sample must be triggerable through a reproducible file, command-line, or network input; compile into a working native executable; preserve the vulnerable behavior after compilation; and retain enough recovered binary structure to independently validate source-to-sink claims. The final benchmark contains 20 samples from 10 projects across 14 commits (Table~\ref{tab:statistic}). For each vulnerability, we reproduce the public PoC in unified Docker, collect failure evidence with AddressSanitizer and Valgrind~\cite{valgrind}, and annotate the attacker-controlled source, vulnerability-relevant sink, triggering propagation chain, and runtime failure evidence. Each annotation is produced by one author and independently checked by another against source code, binary artifacts, crash reports, and CVE descriptions. Constructing the benchmark required over 40 person-days of expert effort, constraining the current scale and motivating reusable, high-fidelity evaluation artifacts. \model{} and all binary-analysis baselines run on stripped binaries, source-code baselines on the vulnerable commit. We compile each project with \texttt{-O0} and strip symbols to preserve binary-level semantic structure under controlled conditions; optimized binaries are discussed in Section~\ref{sec:threats}.

\begin{table}[!t]
\centering
\caption{Statistics of the evaluated projects. \#IR Inst. and \#Func report the size of LLVM IR instructions and source code functions of the analyzed binary; \#Vuln\_files is the number of source files in the vulnerability trace, and \#Vuln is the number of ground-truth vulnerabilities in each commit.}
\label{tab:statistic}
\renewcommand{\arraystretch}{0.9}
\resizebox{\linewidth}{!}{
\begin{tabular}{llllrrrr} 
\toprule
 \textbf{ID} & \textbf{Project}                                     & \textbf{Commit}  & \textbf{Binary} &  \textbf{\#IR Inst.}& \textbf{\#Func} & \textbf{\#Vuln\_files} & \textbf{\#Vuln}  \\
\midrule
\multirow{2}{*}{P1} & matio & 64f7936 &  matdump   & 36,075 &330  &  3                             &     2    \\
& matio & 55e506b &  matdump & 37,588 & 334 &     4                         &     3  \\
\multirow{2}{*}{P2} & libexif & a918830 & exif & 27,737 &385 &      2                         &    1     \\
& libexif & d0ebd6e &  exif & 19,988 &380 &    3                           &     1    \\
\multirow{2}{*}{P3} & libsndfile & b0d7f5b &   sndfile-info    &  126,238 & 1,608  &     4                  &     1    \\
& libsndfile & 1d928bf &   sndfile-deinterleave    &  125,410 & 1,530   &      5                 &     1    \\
% \multirow{2}{*}{P4} & giflib & c63cc98 &   gif2rgb    &  &132   &      2              &     1    \\
% & giflib & b191fe8 &   gif2rgb  &  &133   &      2                 &     1    \\
P4 & giflib & 52b62de &   gif2rgb  & 11,370 &133   &      2                 &     2    \\
P5 & libheif & fd0c01d &   heif-convert    & 78,396 &2,141  &    5                       &     1    \\
% \multirow{2}{*}{P5} & libheif & fd0c01d &   heif-convert    &  2141  &    5                       &     1    \\
% & libheif & 3824054 &   heif-convert    &  2163  &    3                       &     1    \\
% \multirow{2}{*}{P6} & faad2 & 1073aee &   faad    &  609    &    7                    &     2    \\ 
% & faad2 & f71b5e8 &   faad    &   609   &    6                    &     1    \\ 
P6 & libtiff & 1bdbd03 &   tiffcrop     & 107,603 &1,601    &    1                    &     1    \\ 
% \multirow{2}{*}{P6} & libtiff & 1bdbd03 &   tiffcrop     &  1601    &    1                    &     2    \\ 
% & libtiff & 4c3ddd9 &   tiffcrop     &  1570    &    1                     &     1   \\ 
P7 & jhead & 11e6e87 &   jhead     &  11,305  &65  &    2                     &     2   \\ 
% \multirow{2}{*}{P7} & jhead & 11e6e87 &   jhead     &    65  &    2                     &     2   \\ 
% & jhead & 871e319 &   jhead     &    65  &    3                    &     1   \\ 
P8 & ezxml & dcb1748 &  ezxml    & 6,876  &33   &    1                     &     2  \\ 
% \multirow{2}{*}{P8} & ezxml & dcb1748 &  ezxml    &   33   &    1                     &     3  \\ 
% & ezxml & 9df0645 &  ezxml    &  31    &    1                     &     2   \\ 
\multirow{2}{*}{P9} & libredwg & 5f99814 &   dwg2SVG    & 966,014  &1,781   &    3              &     1   \\ 
& libredwg & 93c2512 &   dwg2dxf   & 4,067,615  &2,757   &    2              &     1   \\ 
P10 & faad2 & 1073aee &   faad    &  77,227  &609  &    1              &     1  \\ 
% \multirow{2}{*}{P10} & bacnet-stack & 11efd69 &   bacnet-npdu    &   10476   &    2              &     2   \\ 
% & bacnet-stack & 0682428 &   bacnet-apdu    &   10402   &   3             &     2   \\ 
\bottomrule
\end{tabular}
}
\end{table}
\begin{table}
\centering
\caption{\diff{Baseline and \model{} settings.}}
\label{tab:baseline}
\renewcommand{\arraystretch}{0.9}
\resizebox{\linewidth}{!}{
\begin{tabular}{lcccccc}
\toprule
& \textbf{Source} & \textbf{Binary} & \textbf{Static} & \textbf{Dynamic} & \textbf{Agentic} & \textbf{Specialization} \\
\midrule
Meta Infer~\cite{meta_infer_2025} & $\checkmark$ & & $\checkmark$ & & & - \\
Semgrep~\cite{semgrep_github} & $\checkmark$ & & $\checkmark$ & & & - \\
cwe\_checker~\cite{fkie_cwe_checker} & & $\checkmark$ & & $\checkmark$ & & - \\
AFL++~\cite{aflplusplus} & $\checkmark$ & $\checkmark$ & & $\checkmark$ & & - \\
\midrule
RepoAudit~\cite{guo2025repoaudit} & $\checkmark$ & & $\checkmark$ & & $\checkmark$ & \fullsym \\
Codex~\cite{openai_codex_security} & $\checkmark$ & & $\checkmark$ & $\checkmark$ & $\checkmark$ & \gensym \\
Claude Code~\cite{anthropic_claude_code_overview} & $\checkmark$ & & $\checkmark$ & $\checkmark$ & $\checkmark$ & \gensym \\
CC-source-skill & $\checkmark$ & & $\checkmark$ & $\checkmark$ & $\checkmark$ & \partsym \\
CC-binary-skill & & $\checkmark$ & $\checkmark$ & $\checkmark$ & $\checkmark$ & \partsym \\
\model{} & & $\checkmark$ & $\checkmark$ & $\checkmark$ & $\checkmark$ & \fullsym \\
\bottomrule
\end{tabular}
}
\par\vspace{1mm}
\begin{minipage}{\linewidth}
\footnotesize
\diff{\textit{Note.} \gensym{} denotes general-purpose agents, \partsym{} denotes knowledge-augmented agents, and \fullsym{} denotes task-specific vulnerability reasoning frameworks.}
\end{minipage}
\end{table}

\begin{table*}
\centering
\caption{Ground-truth vulnerability detection results of Veritas and baseline methods. Each project-method cell reports TP/FN over the curated vulnerabilities in that project. The final rows report aggregate TP/FN and overall recall across all evaluated projects. ``N/A'' indicates that the method produced no applicable result for the target vulnerability class.}
\label{tab:detection}
\renewcommand{\arraystretch}{0.9}
\resizebox{\linewidth}{!}{
\begin{tabular}{cccccccccccc} 
\toprule
\multirow{2}{*}{ID} & \multirow{2}{*}{\#Vuln} & Meta Infer~\cite{meta_infer_2025} & Semgrep~\cite{semgrep_rules_0xdea} & cwe\_checker~\cite{fkie_cwe_checker} & AFL++~\cite{aflplusplus} & RepoAudit~\cite{guo2025repoaudit} & Codex~\cite{openai_codex_security} & Claude Code~\cite{anthropic_claude_code_overview} & CC-source-skill & CC-binary-skill & \model  \\ 
\cmidrule{3-12}
                    &                         & TP/FN       & TP/FN    & TP/FN         & TP/FN  & TP/FN      & TP/FN  & TP/FN        & TP/FN & TP/FN & TP/FN  \\ 
\cmidrule{1-12}
P1                  & 5                       &     0/5        & 0/5        & 0/5             & 1/4     & 5/0          &    1/4    & 1/4    & 1/4    &   0/5 &   3/2     \\
P2                  & 2                       &    0/2         & 0/2        & 0/2             & 1/1      & 0/2          &   1/1     & 1/1       &  1/1   &   0/2       &    2/0    \\
P3                  & 2                       &     0/2        & 0/2        & 0/2             &   0/2     & 0/2         &      0/2    &      0/2      &  0/2   &   1/1    &     2/0   \\
P4                  & 2                       &    0/2         & 0/2        & 0/2             & 0/2      & 0/2          & 2/0      & 2/0       & 2/0    & 2/0         &    2/0    \\
P5                  & 1                       &    0/1         & 0/1        & N/A             & 0/1      & 0/1          & 0/1     & 0/1       &  0/1   &   1/0       &   1/0     \\
P6                  & 1                       &    0/1         & 0/1        & 0/1            &     0/1   & 0/1          &     0/1   &        0/1      & 0/1    & 0/1    &   1/0     \\
P7                  & 2                       &   0/2          & 0/2        & 0/2            & 0/2    & 0/2          &   0/2      & 0/2         &  0/2   &     0/2   &     2/0   \\
P8                  & 2                       &     0/2        & 0/2        & 0/2     & 0/2   & 0/2          &    0/2    &      1/1       &  0/2   &   0/2   &    2/0    \\
P9                  & 2                       &   0/2          & 0/2        & N/A              & 0/2     & 0/2          &   0/2      & 1/1       &  0/2   &     1/1     &    2/0    \\
P10                 & 1                      &   0/1          &    0/1      &     0/1           &   0/1      &     1/0       &      1/0  &        1/0      &  1/0   &  0/1   &    1/0    \\ 
\midrule
Total TP/FN         & 20                      &    0/20         & 0/20       & 0/20             &    2/18   &  6/14       &    5/15    &       7/13     & 5/15    &   5/15    &     18/2  \\
Recall (\%)              & -                       &   0.00          &    0.00      &       0.00        & 10.00       &      30.00      &    25.00    &      35.00       &  25.00   &  25.00    &    90.00    \\
\bottomrule
\end{tabular}
}
\end{table*}
\begin{table*}
\centering
\caption{False Positives (FPs) in the Exhaustively Validated Subset.}
\label{tab:fp_exhaustive}
\renewcommand{\arraystretch}{0.9}
\resizebox{\linewidth}{!}{
\begin{tabular}{ccccccccccccc} 
\toprule
Project    & Commit     & \#Vuln       & Meta Infer~\cite{meta_infer_2025} & Semgrep~\cite{semgrep_rules_0xdea} & cwe\_checker~\cite{fkie_cwe_checker} & AFL++~\cite{aflplusplus} & RepoAudit~\cite{guo2025repoaudit} & Codex~\cite{openai_codex_security} & Claude Code~\cite{anthropic_claude_code_overview} & CC-source-skill & CC-binary-skill & \model \\ 
\midrule
libredwg   & 5f99814      & 1     &     144       &    15     &        N/A      &      0 &      2     &   3    &     8   & 12  & 1  &    0   \\
libexif    & d0ebd6e     & 1      &       0     &     4    &      0        &     0  &     0      &    4   &  12        & 6  &   2  &   0    \\
libsndfile & b0d7f5b      & 1     &      41      &      8   &      30        &    0   &       5    &   3    &       7    & 11 &   1 &    0   \\
libheif    & fd0c01d      & 1     &     28       &     2    &       N/A       &   0    &      0     &    3   &      16    & 19 &   1  &    0   \\
giflib     & 52b62de     & 2      &    7        &     3    &        10      &    0   &       13    &    2   &  131         & 18 &   1 &   0    \\
jhead      & 11e6e87    & 2       &      10      &     21    &         2     &   0    &    1       &   2    &      5      & 8 & 4  &    0   \\
faad2      & 1073aee   & 1        &      69      &     26    &          51    &    0   &      0    &   7   &      7    & 3 &   0  &   0    \\ 
\midrule
\multicolumn{2}{c}{Total FP}   & 9 & 299          &  77       &      93        &    0  &      21     &    24   &     186  &  77 & 10  &    0   \\
\bottomrule
\end{tabular}
}
\end{table*}

\subsection{Baseline}
\subsubsection{Baseline Selection}
\label{sec:baselineselection}
\diff{We compare \model{} with representative methods spanning source code analysis, binary analysis, dynamic testing, and agent-driven auditing, summarized in Table~\ref{tab:baseline}. Source-code baselines are included as strong reference points, even though \model{} operates only on stripped binaries.}

\diff{For static analysis, we include three analyzers: \textit{Meta Infer}~\cite{meta_infer_2025}, deep inter-procedural reasoning based on separation logic; \textit{Semgrep}~\cite{semgrep_github}, lightweight rule-based AST matching, which we strengthen on out-of-bounds bugs with expert rules~\cite{semgrep_rules_0xdea}; and \textit{cwe\_checker}~\cite{fkie_cwe_checker}, a binary scanner using data-flow analysis and symbolic execution over Ghidra~\cite{ghidra}.}

\diff{For dynamic testing, we adopt \textit{AFL++}~\cite{aflplusplus}, a standard strategy for memory-safety checking through concrete execution. All targets are recompiled with AFL++ instrumentation and sanitizers; each sample is fuzzed for 8 hours on 3 parallel cores with 5 initial seeds in Docker environments following cost-bounded fuzzing baselines~\cite{klees2018evaluating,bohme2017directed}.}

\diff{For agent-driven auditing, we include RepoAudit~\cite{guo2025repoaudit}, a specialist source-auditing agent; two general coding agents, Codex~\cite{openai_codex_security} and Claude Code~\cite{anthropic_claude_code_overview}; and two skill-augmented Claude Code variants. \textit{RepoAudit} is a multi-agent auditing system with on-demand inter-procedural path construction using GPT-5.4~\cite{openai_gpt54_2026}. To keep the comparison cost-bounded, we scope the reasoning space to files containing vulnerability chains, following the BugScope~\cite{guo2025bugscope} setting. We report this as favorable to RepoAudit, not as an unconstrained repository audit. \textit{Codex} (GPT-5.4) and \textit{Claude Code} (Opus 4.6) are given identical repository access, build/test permissions, and a two-hour audit budget. \textit{CC-source-skill} augments Claude Code with Trail of Bits'~\cite{trailofbits_skills} c-review and fp-review skills. \textit{CC-binary-skill} gives Claude Code access to the same binary-analysis tools and stripped binaries used by \model{} and a customized binary OOB vulnerability analysis skill we developed.}

\subsubsection{Evaluation Protocol}
\diff{We evaluate all methods against the curated flow-level ground truth. For source-code methods, including static and agent-based approaches, Cursor~\cite{cursor2026} (GPT-5.4~\cite{openai_gpt54_2026}) assists matching reported findings to benchmark vulnerabilities, and four security experts review all matches. For binary-based methods, an automated mapping pipeline recovers binary base addresses, maps reports to the ELF layout, and resolves source-level function information with \texttt{addr2line}. A ground-truth vulnerability is a TP if at least one report matches its annotation, else an FN. Duplicate reports for the same vulnerability are counted once. We report TP, FN, and recall, aggregating results by project ID when multiple vulnerable commits share the same project.}

\diff{Findings that do not match any ground-truth vulnerability are counted as false positives (FPs). For \model, FP accounting follows its two-stage pipeline: \textit{Discover} emits candidates, and \textit{Validator} confirms or rejects them with runtime evidence. Because validation cost scales with candidate volume, we exhaustively validate all candidates for 7 project commits covering 9 samples, and sample larger candidate sets for the remaining projects. Each raw \textit{Discover} output is treated as one non-deduplicated validation task. Thus, TP, FN, and recall are computed over the full benchmark, while FPs outside the exhaustive subset are reported as sampled audit estimates over validated candidates rather than exact whole-population FPR.}

\begin{table}
\centering
\caption{\diff{Candidate-level cost comparison.}}
\label{tab:candidate}
\renewcommand{\arraystretch}{0.9}
\resizebox{0.9\linewidth}{!}{
\begin{tabular}{lrrr} 
\toprule
                & \multicolumn{1}{l}{Time / cand. (s)} & \multicolumn{1}{l}{Cost / cand. (\$)} & \multicolumn{1}{l}{Avg. \#Cand.}  \\ 
\midrule
RepoAudit~\cite{repoaudit_bugreports}       & 321.05  &    0.46     &     24.79                \\
Codex~\cite{openai_codex_security}         &  1,076.94    &   5.09     &    5.43               \\
Claude Code~\cite{anthropic_claude_code_overview}     &    106.85   &       0.77        &     26.64      \\
CC-source-skill  &   199.61 &   0.97      &     27.14        \\
CC-binary-skill &   246.47    &     1.27    &         7.43       \\
\midrule
\model\_discover &    30.30     &    0.30                   &     \multirow{2}{*}{188.07}     \\
\model\_validator &   275.97   &  1.79  &                                    \\
\bottomrule
\end{tabular}
}
\end{table}

\subsection{RQ1: Effectiveness of Grounded Vulnerability Reasoning}
We apply \model{} and all baselines to the curated benchmark. Table~\ref{tab:detection} reports TP/FN and recall over ground-truth vulnerabilities, and Table~\ref{tab:fp_exhaustive} reports false positives on the exhaustively validated subset. Traditional tools perform poorly for different reasons. Meta Infer~\cite{meta_infer_2025}, Semgrep~\cite{semgrep_rules_0xdea}, and cwe\_checker~\cite{fkie_cwe_checker} achieve zero recall because they depend on predefined rules, CWE patterns, or conservative static approximations that do not capture the specific propagation and bound relationships in our cases. AFL++ reaches 10\% recall with zero false positives, showing that concrete execution provides reliable evidence when a crash is reached, but unguided mutation rarely reaches the required path and memory state within the bounded fuzzing budget. Agentic baselines improve over traditional tools but remain limited in different ways. RepoAudit~\cite{guo2025repoaudit} detects 6 of 20 vulnerabilities under the favorable file-scoped setting, but still misses propagation through shared state and data structures beyond direct call chains. Codex and Claude Code reach 25\% and 35\% recall, with Claude Code producing many false positives, reflecting the cost of broad, weakly directed exploration. Both skill-augmented variants reduce false positives but remain at 25\% recall: CC-source-skill's general C-review and false-positive skills are not targeted to memory-specific reasoning over allocation, bounds, and unsafe accesses, while CC-binary-skill is memory-oriented but struggles to search fragmented binary artifacts without explicit source-to-sink guidance. This last gap directly motivates the \textit{Slicer}, which supplies witness-backed flows before LLM reasoning.
\model{} achieves 90\% recall, detects 18 of 20 vulnerabilities, outperforms all baselines, and yields only 2 observed false positives across exhaustive and sampled validation (Section~\ref{sec:rq3}). This advantage comes from recovering witness-backed flows, constructing fine-grained source-to-sink claims, and applying executable validation.

\subsection{RQ2: Cost Analysis}
\label{sec:cost}
\diff{RQ2 analyzes candidate-level cost, where a candidate is an intermediate vulnerability hypothesis later validated against executable evidence/false positive check. End-to-end cost conflates per-claim effort with candidate volume, obscuring whether cost comes from expensive reasoning or from preserving more hypotheses for validation. We report time and API cost per candidate in Table~\ref{tab:candidate}, together with the average number of candidates, to separate per-claim efficiency from recall-oriented candidate generation.}

\begin{table}
\centering
\caption{Effect of \textit{Slicer} context construction. Sink Cov., Flow Cov., Func Con., and Avg. \#Funcs denote sink coverage, full-trace coverage, trace connectivity, and average context size.}
\renewcommand{\arraystretch}{0.9}
\resizebox{0.9\linewidth}{!}{
\label{tab:slice_ablation}
\begin{tabular}{lrrrr}
\toprule
Context & Sink Cov. & Flow Cov. & Flow Con. & Avg. \#Funcs \\
\midrule
% CG-1    & 4/21  & 3/21  & 3/21  & 7.2   \\
% CG-2    & 14/21 & 11/21 & 6/21  & 132.8 \\
% CG-3    & 18/21 & 17/21 & 6/21  & 272.6 \\
% CG-4    & 21/21 & 21/21 & 6/21  & 436.9 \\
% CG-5    & 21/21 & 21/21 & 6/21  & 494.4 \\
% \textit{Slicer} & 21/21 & 21/21 & 21/21 & 5.16  \\
CG-1    &     4/20     &        3/20     &  3/20   &     7.2          \\
CG-2    &     14/20     &        11/20      &   6/20 &   136.4            \\
CG-3    &     18/20       &        17/20     &   6/20  &  269.1             \\
CG-4    &      20/20      &      20/20      &    6/20  &    408.9           \\
CG-5    &     20/20       &    20/20      &    6/20   &    446.1           \\
\textit{Slicer} &      20/20      &    20/20       &    20/20   &      5.16        \\
\bottomrule
\end{tabular}
}
\end{table}
\begin{table}
\centering
\caption{\textit{Discover} on the exhaustive subset. \#Flows are witness-backed slicer outputs, \#Cand. are \textit{Discover} candidates, and TP/FN are detected and missed ground-truth vulnerabilities.}
\label{tab:abla_detection}
\renewcommand{\arraystretch}{0.9}
\resizebox{0.7\linewidth}{!}{
\begin{tabular}{llrrr} 
\toprule
Project    & Commit         & \#Flows & \#Cand. & TP/FN  \\ 
\midrule
libredwg   & 5f99814        & 451             & 144               & 1/0    \\
libexif    & d0ebd6e        & 155             & 70                & 1/0    \\
libsndfile & b0d7f5b        & 1,069           & 313               & 1/0    \\
libheif    & fd0c01d        & 146             & 47                & 1/0    \\
giflib     & 52b62de        & 27              & 12                 & 2/0    \\
jhead      & 11e6e87        & 47              & 15                & 2/0    \\
faad2      & 1073aee        & 80              & 22                & 1/0    \\ 
\midrule
\multicolumn{2}{l}{Total}   & 2,002           & 623               & 9/0    \\
\bottomrule
\end{tabular}
}
\end{table}

\diff{Within \model{}, claim discovery is lightweight: \model{}-Discover costs 30.30 seconds and \$0.30 per candidate, since the \textit{Slicer} fixes the flow-level reasoning scope and trie-based memoization reuses shared prefixes. By contrast, Codex has the highest per-claim cost and lowest number of claims, suggesting that much of its budget goes to broad exploration, context inspection, and build/test attempts rather than grounded claim generation. The main cost is \model{}-Validator at 275.97 seconds and \$1.79 per candidate, reflecting the cost of mapping each claim to executable coordinates and checking it under runtime oracles on stripped binaries. The larger number of \model{} candidates reflects fine-grained claim construction. \textit{Discover} preserves source-to-sink claims with explicit propagation, bound, and sink conditions, allowing \textit{Validator} to check each hypothesis against executable evidence. This granularity is important in stripped binaries, where aggressive pruning is risky under incomplete semantics.}

\subsection{RQ3: Component Effectiveness}
\label{sec:rq3}
In this section, we examine the contributions of \model's three stages.
The \textit{Slicer} constructs compact yet semantically complete evidence flows from stripped binaries. We compare its context construction with call-graph expansion under hop budgets from CG-1 to CG-5, measuring sink coverage, full-trace coverage, correct propagation connectivity, and average context size over 20 ground-truth vulnerabilities. Table~\ref{tab:slice_ablation} shows that call-graph expansion requires 4 hops to cover all sinks. Even at CG-4 and CG-5, however, only 6 of 20 cases achieve correct propagation connectivity, because call-chain traversal cannot capture shared-state dependencies that bypass direct caller-callee edges. This coverage comes at high context cost: CG-4 averages 408.9 functions per flow. In contrast, the \textit{Slicer} achieves 20/20 on all three coverage metrics with only 5.16 functions per flow, confirming that grounding context construction in source-to-sink propagation evidence yields higher semantic completeness and substantially smaller inputs.

\textit{Discover} turns this grounded search space into a tractable candidate set without losing ground-truth vulnerabilities. On the exhaustive subset, the \textit{Slicer} emits 2,002 witness-backed flows across 7 project-commits, and \textit{Discover} narrows them to 623 candidates, a 68.88\% reduction, while retaining all 9 ground-truth vulnerabilities (Table~\ref{tab:abla_detection}). This reduction is essential for runtime grounding: without it, the \textit{Validator} would face over three times as many hypotheses under an already expensive per-candidate budget (Section~\ref{sec:cost}).

Table~\ref{tab:fp_large} evaluates the \textit{Validator}'s false-positive filtering under non-deduplicated accounting: across exhaustive and sampled validation, \model\ produces only 2 false positives among 957 validated candidates, versus 10-617 from baselines on the same subset (Table~\ref{tab:fp_exhaustive}). Together, the three ablations show each stage is necessary: the \textit{Slicer} makes context tractable, \textit{Discover} prevents candidate inflation beyond the validation budget, and the \textit{Validator} filters infeasible hypotheses.

\begin{table}
\centering
\caption{False positive analysis of \model\ under non-deduplicated setting. \#Cand. and \#Validated are raw and validated candidates; Coverage is the validated fraction.}
\label{tab:fp_large}
\renewcommand{\arraystretch}{0.9}
\resizebox{\linewidth}{!}{
\begin{tabular}{lrrrrr} 
\toprule
Subset 
& \#Vulns 
& \begin{tabular}[c]{@{}r@{}}\#Cand.\end{tabular} 
& \begin{tabular}[c]{@{}r@{}}\#Validated\end{tabular} 
& \begin{tabular}[c]{@{}r@{}}Coverage\end{tabular} 
& FP \\ 
\midrule
Exhaustive Validation & 9  & 623   & 623 & 100\%   & 0 \\
Sampled Validation    & 11 & 3,344 & 334 & 10\%    & 2 \\ 
\midrule
Total                 & 20 & 3,967 & 957 & 24.12\% & 2 \\
\bottomrule
\end{tabular}
}
\end{table}

\begin{figure}[!t]
    \centering
    \includegraphics[width=0.85\linewidth]{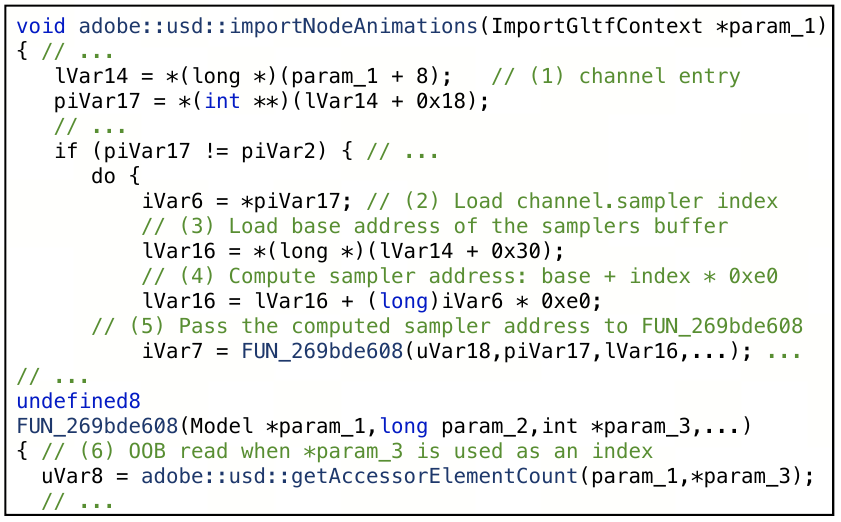}
    \caption{The simplified code of CVE-2025-[REDACTED].}
    \label{fig:case}
\end{figure}

\subsection{Real-world Application}
Applying \model\ to Apple macOS, we discovered a zero-day out-of-bounds read (CVE-2025-[REDACTED]) in \texttt{libusd\_ms.dylib}, Apple's Universal Scene Description library for \texttt{glTF} parsing (Figure~\ref{fig:case}). The bug arises because \texttt{glTF} does not enforce that a channel's \texttt{sampler} index refers to a valid entry in the \texttt{samplers} array. During import, \texttt{importNodeAnimations} loads the attacker-controlled \texttt{channel.sampler} index (2) and scales it by the sampler-object size to index the \texttt{samplers} buffer (4), without any bounds check. When the index is out of range, the computed pointer lies outside the buffer; it is passed to \texttt{FUN\_269bde608} and dereferenced at \texttt{getAccessorElementCount} (6), triggering the out-of-bounds read. This case exercises all three stages of \model\ on a real target: 
Starting from \texttt{UsdGltfFileFormat::Read}, the file-loading entry point, the \textit{Slicer} extracts the \texttt{glTF} flow, tracking the attacker-controlled sampler index through \texttt{importNodeAnimations} into \texttt{FUN\_269bde608}. Dual-view reasoning at the source and sink traces this index from the \texttt{glTF} input through the pointer arithmetic at $(4)$ to the dereference at $(6)$, relating the decompiled offset expression to the lifted-IR memory operation and identifying the missing bounds check. To validate this claim, the \textit{Validator} constructs a proof-of-concept \texttt{glTF} file with an out-of-range sampler index, e.g., \texttt{0x1000000}, and executes it under a debugger with a breakpoint at $(4)$, confirming that the computed pointer exceeds the buffer boundary and triggers a crash.

\section{Related Work}
\paragraph{Out-of-Bounds Vulnerability Analysis}
Out-of-bounds detection has long relied on static and dynamic analysis. Static techniques, from range analysis to abstract interpretation~\cite{wagner2000first,ganapathy2003buffer,zhang2025statically}, reason about bounds broadly but over-approximate, yielding infeasible paths and false positives~\cite{engler2001bugs}. Dynamic techniques such as fuzzing, sanitizers, and concolic execution~\cite{serebryany2012addresssanitizer,jiang2024fuzzing,cadar2008klee,avgerinos2014automatic} provide concrete evidence once a vulnerable path is reached, but struggle with path explosion and deep states~\cite{vadayath2022arbiter}. For binaries, frameworks such as BAP and angr~\cite{brumley2011bap,wang2017angr} lift executables for symbolic reasoning, while detectors such as cwe\_checker and BinAbsInspector~\cite{fkie_cwe_checker,Keen2023BinAbsInspector} encode common memory-safety patterns. Yet fixed predicates struggle with diverse out-of-bounds traces across functions and lossy binary views, where input provenance, object bounds, and unsafe accesses are hard to connect and validate.

\paragraph{LLM/Agent-based Vulnerability Reasoning}
\diff{Early LLM-based approaches identify vulnerability patterns through fine-tuning~\cite{sheng2025llms,ding2024vulnerability,wen2025boosting}, but depend on training-data quality and transfer poorly to real-world repositories~\cite{risse2025top}. Recent systems improve context selection and reasoning control through program structure, retrieval, call-graph expansion, and tool-constrained agent workflows~\cite{lekssays2025llmxcpg,nie2025vulnllm,guo2025bugscope,hussain2025vulbinllm,yildiz2025benchmarking,shao2025craken,abramovichenigma,li2024iris,guo2025repoaudit}, but mainly target source or repository artifacts. For binaries, LATTE~\cite{liu2023harnessing} and VulBinLLM~\cite{hussain2025vulbinllm} reason over decompiled code, but their path construction is not provenance-preserving: LATTE follows call-graph-guided context, while VulBinLLM relies on model-driven retrieval. Code-Augur~\cite{luo2026code} instead falsifies LLM-inferred source invariants with guided fuzzing, but such specifications are fragile when symbols, types, and source structure are unavailable.}

\diff{Frontier models are also improving rapidly, with industrial reports of real vulnerability discoveries in production software~\cite{openai_gpt54_2026,anthropic2026mythos,anthropic2026mozilla,anthropic2026glasswing,fort2026system}. Rather than making semantic grounding unnecessary, these results underscore that the surrounding architecture matters: how context is selected, hypotheses prioritized, and findings validated. This aligns with evidence that unconstrained reasoning remains brittle on subtle semantic distinctions in large contexts~\cite{ding2024vulnerability,yildiz2025benchmarking,steenhoek2024err}. \model\ instantiates this principle for stripped binaries, grounding claim construction in deterministic program-analysis evidence and confirming hypotheses through debugger-visible artifacts.}

\section{Discussion}
\label{sec:discussion}
% \textbf{Dataset Scope}
% Our evaluation uses 20 vulnerability instances from 10 projects. This scale is small relative to source-code benchmarks but reflects genuine constraints of binary-level evaluation: each sample must satisfy four criteria simultaneously (Section~\ref{sec:dataset}), and fine-grained ground-truth annotation demands roughly two person-days per case, however, such annotated binary vulnerability datasets remain scarce~\cite{juliet-ccpp-13,lee2025sec,mei2024arvo}.
% and existing benchmarks either provide coarse labels over synthetic code or lack trace-level flow annotations.

%\paragraph{Hybrid Grounding and Model Evolution}
\model's separation of deterministic analysis, LLM reasoning, and runtime validation reflects current engineering and assurance constraints, not a claim that semantic recovery must remain outside learned models. As frontier models improve~\cite{guo2025bugscope,liu2023harnessing,anthropic2026glasswing,openai_gpt54_2026,fort2026system}, some tasks may migrate into learned components, but the core requirements remain: selecting relevant context, preserving semantic evidence, and validating against execution. We therefore view semantic grounding as an architectural principle, not a fixed implementation boundary.

%\textbf{Complementarity with Fuzzing.}
\model\ is complementary to fuzzing: rather than searching broadly for crashing inputs, it constructs grounded hypotheses from static binary evidence and confirms them through runtime execution. Combining witness-backed candidates with directed fuzzing or symbolic execution is a natural extension.
\section{Threats to Validity}
\label{sec:threats}
\diff{\textbf{External validity.}
We instantiate \model{} on out-of-bounds vulnerabilities because they are prevalent and match the evidence structure studied here: source-to-sink propagation, object-bound reasoning, and runtime-oracle confirmation. Extending \model{} to other classes, such as use-after-free or double-free, requires bug-specific grounding semantics and runtime oracles. We also compile targets with \texttt{-O0} to study semantic grounding under controlled binary-recovery conditions; optimized binaries remain future work.
% We instantiate \model{} on out-of-bounds vulnerabilities because they are prevalent and naturally match the evidence structure studied here: source-to-sink propagation, object-bound reasoning, and runtime-oracle confirmation. Extending \model{} to other classes, such as use-after-free or double-free, would require bug-specific grounding semantics, including lifetime-related sources and sinks, allocation/release/use evidence, and runtime oracles for temporal-safety violations. Meanwhile, we compile targets with \texttt{-O0} to study semantic grounding under controlled binary-recovery conditions; optimized binaries with inlining and transformed control flow remain future work.
}

\diff{\textbf{Dataset and construct validity.}
Our evaluation covers 20 vulnerability instances from 10 projects. This scale is smaller than source-code benchmarks but reflects the cost of binary-level evaluation, where each case must compile, preserve vulnerable behavior after stripping, and support verifiable flow-level ground truth. This protocol enables claim-level evaluation, but may bias the benchmark toward vulnerabilities whose provenance remains recoverable from binaries.
% Our evaluation covers 20 vulnerability instances from 10 projects. Although smaller than source-code benchmarks, this scale reflects the cost of binary-level evaluation: each case must compile, preserve the vulnerable behavior after stripping, and support manually verifiable flow-level ground truth. This protocol enables claim-level evaluation, but may bias the benchmark toward vulnerabilities whose provenance remains recoverable from binary artifacts.
}

\diff{\textbf{Implementation and cost.}
\model\ relies on RetDec-lifted LLVM IR and decompiled code. Severe lifting or decompilation failures may remove evidence needed by \textit{Slicer} or mislead \textit{Discover}, although the modular design allows alternative backends. Runtime validation dominates cost; the reported figures reflect exhaustive research evaluation rather than optimized deployment, and prioritization, deduplication, parallel validation, and instrumentation reuse could reduce it.}

% \diff{\textbf{Internal validity.}
% % \model{} relies on RetDec-lifted LLVM IR and decompiled code, so severe lifting or decompilation failures may remove evidence needed by \textit{Slicer} or mislead \textit{Discover}. Since the grounding layers are modular, the lifting and decompilation backends can be replaced without changing the overall architecture.
% \model{} relies on RetDec-lifted LLVM IR and decompiled code. Severe lifting or decompilation failures may remove evidence needed by \textit{Slicer} or mislead \textit{Discover}. The grounding architecture is modular, so the lifting and decompilation backends can be replaced.
% }

% Runtime validation dominates cost because candidate claims must be mapped to executable coordinates and checked through instrumentation, debugging, and runtime oracles. The reported cost reflects exhaustive candidate-level characterization under a research protocol, not optimized deployment. Deployment could reduce cost through prioritization, deduplication, early stopping, parallel validation, and instrumentation reuse.}
% Runtime validation dominates cost because candidate claims must be mapped to executable coordinates and checked by repeatedly running, instrumenting, and debugging binaries. The reported aggregate validation time should therefore be read as the cost of exhaustive candidate-level characterization under a research protocol, not as optimized deployment cost. Practical deployments could reduce this cost through prioritization, deduplication, early stopping, parallel validation, and instrumentation reuse.}

\section{Conclusion}
%We presented \model, a framework for detecting memory-corruption vulnerabilities in stripped binaries, grounded in two layers: static evidence that constrains LLM reasoning to witness-backed propagation flows from lifted LLVM IR, and runtime evidence that confirms conclusions through concrete execution artifacts. \model\ implements this through a deterministic static slicer, a dual-view step-wise LLM detector, and a multi-agent runtime validator. On 20 real-world binary vulnerability cases, \model\ achieves 90\% recall with few false positives in exhaustively validated runs, outperforming all baselines, and discovered a confirmed Apple zero-day. We believe context selection grounded in program analysis evidence and executable validation will remain core requirements for reliable binary vulnerability discovery.

We presented \model, a framework for grounded vulnerability reasoning over stripped binaries. \model\ combines static grounding, which constrains LLM reasoning to witness-backed source-to-sink flows recovered from lifted LLVM IR, with runtime grounding, which confirms candidate claims against executable behavior. Instantiated for out-of-bounds vulnerabilities, \model\ integrates a deterministic \textit{Slicer}, a dual-view step-wise \textit{Discover}, and a multi-agent \textit{Validator}. On 20 real-world cases, it achieves 90\% recall, produces few false positives, outperforms all baselines, and discovers a confirmed Apple zero-day. These results show that reliable binary vulnerability discovery benefits from deterministic evidence recovery before claim construction and executable confirmation afterward.

\bibliographystyle{IEEEtran}
\bibliography{reference}


\appendix

\section{Appendix}
\subsection{Prompt Design}
Prompting in \model\ is tailored to each stage rather than handled by a single template. The \textit{Detector} uses structured templates for each function and flow history to maintain taint, bounds, and control flow context along each candidate flow. The \textit{Validator} uses role-specific instructions and constrained tool interfaces to keep planning, execution, and failure diagnosis auditable. Thus, prompting reflects the distinct roles of the two LLM-facing components: semantic interpretation over precomputed flows and executable validation of concrete hypotheses.

\subsection{Algorithm of \textit{Detector}}
Algorithm~\ref{alg:detector} summarizes the Dual-view Vulnerability Detector. Given a flow object $\pi=((f_1,\ell_1),\ldots,(f_N,\ell_N))$ with per-function grounded labels, anchor cues $\mathcal{A}(f_n)$, decompiled code $D(f_n)$, lifted IR $I(f_n)$, and a prefix cache $\mathcal{C}$, the \textit{Detector} first retrieves the longest cached prefix (Line~1). If no prefix is cached, the reasoning state is initialized as empty (Line~2). It then processes the remaining functions in flow order (Lines~3-11). Source and sink functions use a dual-view representation combining lifted IR and decompiled code (Line~5), while intermediate functions use only decompiled code (Line~7). At each step, the LLM $\Phi_\theta$ updates the accumulated reasoning state $S_n$ from the previous state, the current function representation, and its anchor cues (Line~9); the resulting state is stored in the prefix cache for later reuse by flows with the same prefix (Line~10). After all functions are processed, the \textit{Detector} performs path-sensitive verification over the complete flow using the final reasoning state $S_N$ (Line~12), and deduplicates violations corresponding to the same memory-safety issue (Line~13). The flow is reported as vulnerable if any violation remains (Line~14), and benign otherwise (Line~15). The final output is the decision $z_\pi \in \{0,1\}$ with a structured explanation derived from the violation set $\mathcal{V}_\pi$ and the accumulated reasoning state $S_N$ (Line~16).

% \begin{algorithm}[t]
% \caption{Dual-view Vulnerability Detector}
% \label{alg:detector}
% \small
% \KwIn{Flow $\pi=(f_1,\dots,f_N)$; anchors $A(f_n)$; decompiled code $D(f_n)$; lifted IR $I(f_n)$; prefix cache $C$}
% \KwOut{Decision $z_\pi \in \{0,1\}$ and explanation}
% $(\pi^{(n^*)}, S_{n^*}) \leftarrow \texttt{LongestCachedPrefix}(\pi, C)$\;
% \lIf{$n^* = 0$}{$S_0 \leftarrow \emptyset$}
% \For{$n \leftarrow n^*+1$ \KwTo $N$}{
%     \lIf{$A(f_n)\neq\emptyset$}{$\mathcal{R}(f_n) \leftarrow \langle I(f_n), D(f_n)\rangle$}
%     \lElse{$\mathcal{R}(f_n) \leftarrow D(f_n)$}
%     $S_n \leftarrow \Phi_{\theta}(S_{n-1}, \mathcal{R}(f_n), A(f_n))$\;
%     $C[(f_1,\dots,f_n)] \leftarrow S_n$\;
% }
% $\mathcal{J}_{\pi} \leftarrow \texttt{ReflectSelect}(S_1,\dots,S_N)$\;
% $\Omega_{\pi} \leftarrow \emptyset$\;
% \ForEach{$j \in \mathcal{J}_{\pi}$}{
%     $\Omega_{\pi} \leftarrow \Omega_{\pi} \cup \{\texttt{ReflectNoteGeneration}(S_j, f_j)\}$\;
% }
% $z_\pi \leftarrow \Phi_{\theta}(S_N, \Omega_{\pi}, \mathcal{R}(f_N))$\;
% \Return $(z_\pi, \text{explanation})$\;
% \end{algorithm}

% \begin{minipage}{\columnwidth}
\begin{algorithm}
\caption{Dual-view Vulnerability Detector}
\label{alg:detector}
\small
\KwIn{Flow object $\pi=((f_1,\ell_1),...,(f_N,\ell_N))$ with grounded labels $(\ell_1,\dots,\ell_N)$; anchors $\mathcal{A}(f_n)$; decompiled code $D(f_n)$; lifted IR $I(f_n)$; prefix cache $\mathcal{C}$}
\KwOut{Decision $z_\pi \in \{0,1\}$ and explanation}
$(\pi^{(n^*)}, S_{n^*}) \leftarrow \texttt{LongestCachedPrefix}(\pi, \mathcal{C})$\;
\lIf{$n^* = 0$}{$S_0 \leftarrow \emptyset$}
\For{$n \leftarrow n^*+1$ \KwTo $N$}{
    \uIf{$f_n \in \mathcal{F}_{src} \cup \mathcal{F}_{sink}$}{
        $\mathcal{R}(f_n) \leftarrow \langle I(f_n), D(f_n)\rangle$\;
    }
    \Else{
        $\mathcal{R}(f_n) \leftarrow D(f_n)$\;
    }
    $S_n \leftarrow \Phi_{\theta}(S_{n-1}, \mathcal{R}(f_n), \mathcal{A}(f_n))$\;
    $\mathcal{C}[\pi^{(n)}] \leftarrow S_n$\;
}
$\mathcal{V}_\pi \leftarrow \texttt{PathSensitiveVerify}(\pi, S_N, \mathcal{R}(f_N), \mathcal{A}(f_N))$\;
$\mathcal{V}_\pi \leftarrow \texttt{DeduplicateViolations}(\mathcal{V}_\pi)$\;
\lIf{$\mathcal{V}_\pi \neq \emptyset$}{$z_\pi \leftarrow 1$}
\lElse{$z_\pi \leftarrow 0$}
\Return $(z_\pi, \texttt{GenerateExplanation}(\mathcal{V}_\pi,S_N))$\;
\end{algorithm}
% \end{minipage}

\subsection{Failure Case Analysis}
\label{sec:failure_case}
\model\ missed two samples in P1, both share the same root cause: incomplete binary recovery by RetDec at the function preamble. In these samples, part of the local object initialization was absent from both the decompiled C and the lifted LLVM IR. The \textit{Slicer} recovered the correct source-to-sink flows, but the \textit{Detector} could not reconstruct the true extent of the destination object: it identified the correct vulnerability class and missing-boundary condition, but hypothesized a smaller object than actually existed. Under our ground-truth protocol, these are counted as false negatives because the reported triggering condition did not match the actual vulnerability semantics. These false negatives then produced two false positives: the \textit{Validator} confirmed that the program could perform the write length hypothesized by the Detector, but because the vulnerable object was embedded within a larger stack-allocated structure, the write fell within the true bounds and produced neither a crash nor a Valgrind-detectable violation, leaving the \textit{Validator} without a stronger oracle to reject the claim. We classify these as a coupled failure mode caused by representation loss in the recovered binary artifacts: an incorrect boundary hypothesis at detection time propagated through validation, yielding two false positives. 

\end{document}